\documentstyle[twocolumn,aps,psfig]{revtex}

\begin{document}
\draft

\twocolumn[\hsize\textwidth\columnwidth\hsize\csname @twocolumnfalse\endcsname

\title{Effect of Spin-1 Impurities in Dimerized Heisenberg Chains}

\author{P. M. Hansen$^1$, J. A. Riera$^1$, A. Delia$^2$, and E.
        Dagotto$^2$}

\address{$^1$Instituto de F\'{\i}sica Rosario (CONICET) and Departamento
de F\'{\i}sica, \\ Avenida Pellegrini 250, 2000 Rosario, Argentina   \\
$^2$National High Magnetic Field Lab and
Department of Physics, Florida State University,\\ Tallahassee,
Florida 32306,USA}
\maketitle

\begin{abstract}
The effects of spin-1 impurities introduced in dimerized 
antiferromagnetic spin chains are studied using
Lanczos diagonalization and quantum
Monte Carlo simulations. Firstly, the magnitude of the interaction 
between
a Ni-ion replacing a Cu-ion is derived for the case study system 
Cu-O-Cu. Then,
it is shown that the introduction of $S=1$ impurities leads to
the suppression of the original spin gap of the dimerized system
and to the enhancement of the antiferromagnetic correlations close
to the impurity. 
These features are interpreted in terms of a
valence bond picture of the ground state of both the pure and
the doped system. The similitudes and differences between the 
underlying mechanisms that produce these effects in the cases
of nonmagnetic and $S=1$ impurities are analyzed.
It is stressed that a singlet formed between the spin-1 impurity
site and the two spin-1/2 neighbors spins behaves as an effective
nonmagnetic vacancy which may be partially responsible for the 
enhancement of the antiferromagnetic correlations on the whole
system. Implications of these results for experiment are also
discussed.
\end{abstract}


\vskip2pc]
\narrowtext

\section{Introduction}
\label{intro}

One-dimensional or quasi-one-dimensional magnetic systems have several
fascinating properties which continue to attract an intense
theoretical activity. One of these properties is the presence of
a spin gap in antiferromagnetic Heisenberg chains with integer 
spin~\cite{haldane,renard} and with spin-1/2 in
ladder geometries~\cite{ladders,azuma},
as well as in chains with frustrating interactions~\cite{majumdar},
and in the presence of dimerization.
In the latter systems, the dimerization frequently
appears as a consequence of the spin-phonon coupling, i.e. it is
due to a spin-Peierls (SP) instability.
The interest in the spin-Peierls phenomena was recently revived after
the first inorganic SP compound, CuGeO$_3$, was found~\cite{hase}.
A considerable theoretical effort has been devoted to understanding
the properties of this compound~\cite{rieradobry,castilla}. Most of
these studies have been based on a one-dimensional (1D) dimerized
Heisenberg model, i.e. retaining only the spin degrees of freedom and
neglecting the spin-phonon dynamics.
Although most dimerized systems are spin-Peierls systems, there are
also several compounds where the intrinsic structure
leads to dimerized spin chains. Actually
Cu(NO$_3$)$_2\cdot2.5$H$_2$O was the first inorganic alternating-chain
system to be investigated~\cite{bonner}.
More recently, inelastic neutron scattering measurements and 
susceptibility data indicate that the compound CuWO$_4$ consists
of weakly coupled alternating chains~\cite{lake}. 
Another example of structural dimerization is provided by the
CuO$_2$ chains which are present in 
${\rm Sr_{14}Cu_{24}O_{41}}$~\cite{hiroi} which also contains
${\rm Cu_2O_3}$ ladders. In addition,
the compound ${\rm (VO)_2P_2O_7}$, which was previously believed
to be composed by two-leg spin ladders, is
actually better described by alternating spin chains~\cite{garrett}.

The experimental 
introduction of impurities in these magnetic systems has provided
additional information in the quest to
examine the nature of their ground state and
excitations~\cite{ajiro}.
In particular, inorganic systems, such as CuGeO$_3$,
can be easily doped with both non-magnetic and magnetic impurities
i.e. by partially replacing the Cu$^{2+}$ ions by Zn$^{2+}$
and Ni$^{2+}$ ions, respectively. The effect of impurities on CuGeO$_3$
has been investigated using neutron scattering~\cite{lussier,sasago},
magnetic susceptibility measurements~\cite{lussier,hase2},
specific heat measurements~\cite{oseroff}, and
spin electron resonance~\cite{luca}. The main results of these
studies are the collapse of the spin gap~\cite{lussier,hase2}, and
the competition between the SP and the N\'eel
states~\cite{lussier,oseroff}, when a small amount of impurities
is introduced. From the theoretical point of view,
the introduction of non-magnetic impurities, which 
for the nearest-neighbors Heisenberg model leads to the
appearance of chains with open boundary conditions (OBC),
has been analyzed using conformal
field theory and Monte Carlo simulations~\cite{eggert1,eggert2}. 
The main result of these studies is the
appearance of enhanced staggered spin order, which is maximized
near the vacancies. More recently, it has been shown that 
spin-1/2 states are localized near vacancies in dimerized
chains and other systems~\cite{martins2,laukamp,sandvik}.
The experimentally observed stabilization of a N\'eel state
upon Zn doping in dimerized chains can be
related to the enhancement of antiferromagnetic (AF) correlations
near the vacancies which arises
due to the presence of these bound states. The effect has been
studied theoretically  using both analytical~\cite{fukuyama} and
numerical techniques~\cite{martins2,laukamp}.
These bound states are expected to produce spectral weight
inside the original gap~\cite{martins1}, and 
there is already direct experimental evidence compatible with the existence of
 these low-energy excitations~\cite{azuma2,lemmens}. Results very
similar to those obtained in dimerized chains have also been
obtained in ladder systems both experimentally~\cite{ladderexper} and 
theoretically~\cite{martins2,laukamp,martins1,ladderall}.

In the present article, the previous studies performed on
1D AF spin-1/2 dimerized Heisenberg chains~\cite{martins2,laukamp,martins1} 
are extended to
the case of $magnetic$ impurities, in particular spin-1 impurities. In
previous literature, it has been studied the related case of
spin-1 AF Heisenberg chains doped with non-magnetic impurities which
become, again, equivalent to chains with OBC~\cite{s1chains,fabrizio}.
The most important prediction in this context is the presence of
isolated spin-1/2 states on the chain edges.
Experimentally, magnetization and electron spin resonance studies
on the compound ${\rm Ni(C_3H_{10}N_2)_2N_3(ClO_4)}$ 
have confirmed the existence of these
end-chain spin-1/2 spins~\cite{s1exp}. The end-states are easily
explained by the valence bond solid (VBS) picture of the ground
state of spin-1 Haldane chains~\cite{affleck}. 
In the same spirit,
an unifying picture based on a short-range resonant-valence-bond
(RVB) scenario~\cite{anderson,kivelson} has been formulated to
explain the existence of
spin-1/2 bound states in gapped spin-1/2 systems, as
discussed in the previous paragraph~\cite{martins2,laukamp}.

In the present study of spin-1/2 dimerized chains doped with spin-1
impurities, both  RVB and VBS pictures will be used to interpret
our results which were obtained using Lanczos diagonalization and
quantum Monte Carlo (QMC) simulations.
Our main conclusion is that  the overall effects, such as the
destruction of the spin gap by low-energy
excitations due to bound states
near the impurities, are similar to those generated by nonmagnetic
impurities in spite of the fact that the microscopic origin and
details of these bound states are somewhat different. This difference
manifests in  a more localized enhancement of the AF correlations
in the immediate vicinity of 
the $S=1$ impurity than in the case of nonmagnetic
impurities. However, at distances  beyond
one lattice spacing away from the $S=1$ impurity
the spin-spin correlations behave very similarly to the case of doped
nonmagnetic impurities. Thus, globally magnetic and nonmagnetic
impurities apparently lead to similar physics including enhanced
antiferromagnetic correlations.

We wish to emphasize that in this article we are mainly concerned 
with dimerized chains, containing only spin degrees of freedom,
and which can be realized in nature independently
of the spin-Peierls mechanism. The spin-Peierls transitions
recently found in inorganic materials only add an extra motivation
to study dimerized chains since they can describe reasonably
well certain features of these materials at very low
temperatures. To properly describe some other features, the 
inclusion of phononic degrees of freedom is strictly necessary 
at least at 
the mean field level\cite{augier,feiguin,hansen} (see e.g., 
discussion in Section \ref{monte}). In this sense,
our purpose is to give just indications that some
of the results we found could be related to some experimental
findings in spin-Peierls compounds. It is quite likely that
similar experimental studies are going to be carried on in the
near future on other materials containing dimerized chains.

The paper is organized as follows. In Section \ref{effective}
the strength of the effective 
Heisenberg interaction between a Cu-ion and an 
impurity Ni-ion is computed starting from a multiband Hubbard model. In 
Section \ref{lanczos}, the dynamic spin structure factor is studied in 
order to analyze the suppression of the spin gap
in a dimerized chain by the introduction of spin-1 impurities. The
main differences with respect to the case of nonmagnetic
impurities are studied. The spin-spin correlations near
the impurities have also been investigated in order to evaluate the
strength of  the AF order enhancement. In
Section \ref{monte}, results for the magnetic 
susceptibility obtained by quantum Monte Carlo simulations are
presented.  Finally, in the Conclusions section, 
the most important results are discussed, together with their
interpretation within a simple valence bond picture.
Experimental consequences of our calculations are here also discussed.

\section{Effective interaction}
\label{effective}

Before addressing the spin chain model and computational techniques
used in this paper, it is important to estimate the
coupling between the $S=1$ impurity and neighboring
 $S=1/2$ ions. In previous studies based on Zn-doping,
 this calculation was not necessary since Zn$^{2+}$ has 
zero spin. However, for the case of Ni$^{2+}$ impurities with
$S=1$, a Heisenberg coupling between Ni and Cu
should be part of any model that attempts to
describe  their behavior. 
In this section a rough estimation of the
nickel-copper exchange will be made. The calculation will be carried out
for the case of a $180^\circ$ Cu-O-Ni bond, which is relevant for
a variety of materials such as the
1D compounds ${\rm Sr_2 Cu O_3}$ and ${\rm Sr Cu O_2}$, the
Cu-based ladder systems mentioned in the previous section, the High-Tc
cuprates, and several others. The results are intended to provide just
qualitative information about the strength of the exchange. In
principle, they do not apply directly to systems such as CuGeO$_3$
where the Cu-ions are linked by two oxygens with $\sim 90^o$ bonds.
Carrying out a calculation for this particular material would be difficult
because small exchange couplings are expected to emerge from a problem
involving large energy scales of the order of eV, and the rough one-bond
calculation discussed below may not be enough to produce reliable results.

The Hamiltonian analyzed in this section is given
by $H = H_t + H_V + H_H$, with the hopping term 
defined as
\begin{eqnarray}
\nonumber
H_t = &-& \sum_{\sigma} 
       t_{CuO}(c^\dagger_{Cu, \sigma} c_{O, \sigma} + h.c.) \\
      &-& \sum_{\sigma,\alpha} 
       t_{NiO\alpha}(c^\dagger_{Ni \alpha, \sigma} 
            c_{O \alpha, \sigma} + h.c.),
\label{hamhop}
\end{eqnarray}
\noindent 
where $\sigma$ denotes the spin, and $\alpha = 1,2$ the two
active orbitals of Ni (labeled as
Ni1 and Ni2). $c_{Cu,\sigma}$ and $c_{Ni\alpha, \sigma}$
are destruction operators for fermions on the Cu and Ni$\alpha$
orbitals, respectively. $t_{CuO}$ ($t_{NiO\alpha}$) is the hopping
amplitude for fermions to move between Cu and O (Ni$\alpha$ and O).
The rest of the notation is standard. The interaction $H_V$ is defined as 
\begin{eqnarray}
\nonumber
H_V &=& U_{Cu} n_{Cu, \uparrow} n_{Cu, \downarrow} +
      U_O n_{O, \uparrow} n_{O, \downarrow}                \\
\nonumber
    &+& U_{Ni} 
     \sum_{\alpha} n_{Ni\alpha,\uparrow} n_{Ni\alpha,\downarrow}  \\
    &-& \Delta_{Cu} n_{Cu}
      -\Delta_{Ni} \sum_\alpha n_{Ni\alpha},
\label{hamint}
\end{eqnarray}
\noindent 
where $n_{Cu, \sigma}$ is the number operator at the $Cu$ site with
spin $\sigma$, $n_{Cu} = 
n_{Cu, \uparrow} + n_{Cu,\downarrow}$,
and the notation for Ni and O follows  analogous definitions. 
$U_{Cu}, U_{O}, U_{Ni}$ regulate the on-site Coulomb repulsion
for Cu, O, and Ni, respectively. $\Delta_{Cu}$ is the energy
needed to move a fermion
from Cu to O, while $\Delta_{Ni}$ is 
a similar quantity but to move fermions from the two $Ni$ orbitals
(assumed degenerate) to O.
Finally,
\begin{eqnarray}
H_H = - J_H {{{\bf S}_1}\cdot{{\bf S}_2}}
\label{hameff}
\end{eqnarray}
\noindent 
is the ferromagnetic Hund coupling ($J_H > 0$)
between the two active orbitals at the Ni site (with
${\bf S}_{\alpha}$ being the spin at orbital $\alpha$).

This Hamiltonian can be solved exactly for the case of
interest that contains 3 electrons, two of which are expected
to be mostly at the Ni-ion and the other at the Cu-ion. From the 
difference in energy between the ground state and the
first excited state with spin 1 in the spectrum
the exchange coupling can be calculated. The matrix that needs
to be diagonalized is $24 \times 24$.

The parameters corresponding to the Cu and O ions
are approximately known from previous
theoretical and experimental work 
in the context of high-temperature superconductors.
The actual parameter values  used here are~\cite{review}
$\Delta_{Cu} \sim 3.6 eV$, $t_{CuO} \sim 1.3 eV$,
$U_{O} \sim 4.6 eV$,
and $U_{Cu} \sim 10.5 eV$. The rest of the parameters
related with Ni-oxides were discussed in previous
work by two of the present authors and collaborators~\cite{nio},
in the context of hole doped $S=1$ chains.
In that publication it was argued that $\Delta_{Ni} \sim 6.0 eV$,
$U_{Ni} \sim 9.5 eV$, $t_{NiO1} \sim 1.3 eV$, 
$t_{NiO2} \sim 0.75 eV$, and $J_{H} \sim 1.3 eV$, based on a 
variety of experimental work and some cluster calculations.
These will be the parameters used in the present study.

As a warm-up, the similar cases of  $180^\circ$ Cu-O-Cu and Ni-O-Ni bonds
were studied using the same method, 
since the exchanges for CuO$_2$ and NiO$_2$ planes
are known experimentally and, thus, a comparison between the computational
results and properties of the real materials can be carried out. 
The Hamiltonian parameters used are the same
as given before. For the Cu-O-Cu (Ni-O-Ni) bond the 
size of the matrix to diagonalize is $9\times 9$ ($100 
\times 100$) for the case of  two (four) fermions in the problem.
For Ni-O-Ni, four fermions are needed to represent
the two expected spin-1 states in this bond. For the 
copper oxides, the exact diagonalization of the 
Hamiltonian matrix obtained from Eqs.(1-3)
leads to an exchange $J_{Cu-O-Cu} = (100\pm 30) meV$,
while for the nickel oxides it is $J_{Ni-O-Ni} = (20 \pm 10)meV$.
The error bars are crudely  estimated as follows: 1. the many
parameters entering the Hamiltonian were
arbitrarily changed one by one by $\pm 20\%$ (to simulate
uncertainties in their experimental determination), and in each case
the new value of $J$ was calculated, 2. a histogram with the
distribution of $J$'s was built, and from a fit of these results with
a Gaussian function, the error bars were determined as the width of such 
a Gaussian.
It was found that
the parameters that affect the exchange the most are the
hopping amplitudes and $\Delta_{Cu}$. The Coulomb energies and Hund
coupling are
large enough such that small variations in their values do not
affect quantitatively the results.

Considering the roughness of the calculation, it is gratifying that
the theoretically calculated Heisenberg couplings presented above
are in very good agreement with the experimental
results for both materials
(see for instance Ref.\onlinecite{sugai}). Thus, it is expected
that the theoretical prediction 
for $J_{Ni-O-Cu}$ will be at least
qualitatively correct as well. The result obtained in this paper is
$J_{Ni-O-Cu} = (100 \pm 50) meV$ i.e. very similar to 
the case of copper-oxides. Then, it is here estimated that using a 
spin model with the same
exchange $J$ among the $S=1/2$ spins, and among these spins and
the $S=1$ impurity, represents a good approximation to the problem
investigated in this paper.

\section{Dynamical structure factor and spin-spin correlations.}
\label{lanczos}

The one-dimensional model which describes a spin-1/2 Heisenberg
chain in the presence of spin-1 impurities is defined as:
\begin{eqnarray}
{\cal H} &=& J \sum_{i \neq i_0,i_0 -1} (1 + \delta_i) {\bf S}_i \cdot
{\bf S}_{i+1} \nonumber \\
&+& J^\prime \sum_{i_0} \{(1 + \delta_{i_0-1}) {\bf S}_{i_0-1} \cdot
{\bf s}_{i_0} \nonumber \\
&~& ~~~~~~~~~+(1 + \delta_{i_0}) {\bf s}_{i_0} \cdot {\bf S}_{i_0+1} \},
\label{hamtot}
\end{eqnarray}
\noindent 
where ${\bf S}_i$ denote the spin-1/2 degrees of freedom, 
and ${\bf s}_{i_0}$ are
the spin-1 operators located at the impurity sites ${i_0}$. The
dimerization is staggered and it
 corresponds to $\delta_i = (-1)^i \delta$. Periodic
boundary conditions are imposed unless otherwise stated. 
According to the results of the
previous section, in the following the
case $J^\prime = J$ will be considered. 
Note that in Hamiltonian Eq.(\ref{hamtot})
an interaction when two spin-1 impurities
are located next to each other has not been included due
to the result of the previous section suggesting that
$J_{Ni-O-Ni} \ll J_{Cu-O-Cu}, J_{Cu-O-Ni}$.


Using Lanczos
diagonalization and the standard continued fraction
formalism~\cite{review}, the zero temperature spin-spin correlations and
dynamical structure factor for model Eq.(4) have been computed. In
order to analyze finite-temperature properties  a
world-line Monte Carlo algorithm~\cite{WLMC} suitable
for the present problem has been implemented.
This method has allowed us to reach low enough
temperatures such that the
ground state behavior of spin-spin correlations can be calculated. Thus, 
with this technique it is possible
to treat longer chains than those accessible to the
Lanczos method, having only a small finite-temperature contamination.

\subsection{One impurity}

Let us first investigate whether a suppression of the spin gap
occurs when S=1 impurities are introduced in a dimerized chain,
similarly as it occurs in Zn-doped systems. The analysis
here follows closely the study  performed in Ref. \onlinecite{martins1}
for the case of nonmagnetic impurities. Since (i) the spin gap opens at
$k=\pi$, and (ii) the most intense peak of the dynamical structure 
function
$S(k,\omega)$ also appears at $k=\pi$, in agreement with neutron
scattering experiments, then the present analysis starts
by studying the behavior
of $S(\pi,\omega)$ upon the introduction of $S=1$ impurities.

The dynamical structure factor at zero temperature is 
defined by:
\begin{eqnarray}
S^{zz}(k,\omega)=\frac{1}{N} \sum_n |\langle \psi_n | S^{z}(k) |
\psi_0 \rangle |^2 \delta(\omega-(E_n-E_0)),
\label{strucfactor}
\end{eqnarray}
\noindent
where $S^z(k) = (1/N) \sum_m S^z_m
e^{ikm}$, $N$ is the number of sites, $m$ labels the sites of the 
lattice, and the rest of the notation is standard.
In the case of total dimerization, i.e. $\delta = 1$, it can be easily
deduced that the introduction of a single impurity generates a peak
in $S(\pi,\omega)$ inside the gap, actually
 at $\omega=0$. This is due to the fact that the matrix element
$M = \langle \eta_0 | S^{z}(k) | \eta_0 \rangle  = (1/3)(2-(1/2)e^{ik})$
is nonzero. Here $| \eta_0 \rangle$ is
the ground state of the 2-site dimer that carries the impurity i.e.
$| \eta_0 \rangle = \frac{1}{\sqrt{3}} (| 0, \uparrow \rangle -
\sqrt{2} | 1, \downarrow \rangle )$, where $1, 0, -1$ label the states
of the spin-1, and $\uparrow, \downarrow$ correspond to the spin-1/2.
This simple mechanism is also present for impurities of higher spin such
as Co,~\cite{cobalt}
and actually includes Zn as a special case. The important point here
is that in the bond being doped a state with a total nonzero spin appears
after a spin-1/2 is replaced by a spin different from 1/2. 
This nonzero spin in the dimer where the
doping occurs is decoupled from the rest of the lattice
in the extreme limit $\delta = 1$ and, thus, produces weight at zero
energy in the dynamical spin structure factor.


\subsection{Two Impurities}

Let us now consider the effect of two impurities located at arbitrary
distances along a chain. If two impurities are located in a periodic
chain of $N$ sites, it is also easy to understand in the case of
total
dimerization $\delta = 1$ (and remembering that the ground state
is selected to have a total spin projection on the $z$-axis equal to
zero to mimic the case of $\delta < 1$) that if the first impurity 
is located at site 1 then a spectral weight at $\omega =0$ appears 
only if the second impurity is
located at sites $l_2$ which are even. For instance, if $l_2$ is 2
then two free spin-1 are located in the same dimer which naturally 
will produce weight at zero frequency (remember that the Ni-Ni 
exchange has been neglected here since it is much smaller than the 
Cu-Cu or Cu-Ni exchanges). However, if $l_2 = 3$, then the matrix 
element $M$ mentioned in the previous paragraph will have opposite 
signs for the dimers involving
sites 1-2 and 3-4 (note again that the total spin of the global

\vspace{-0.5cm}
\begin{figure}
\psfig{figure=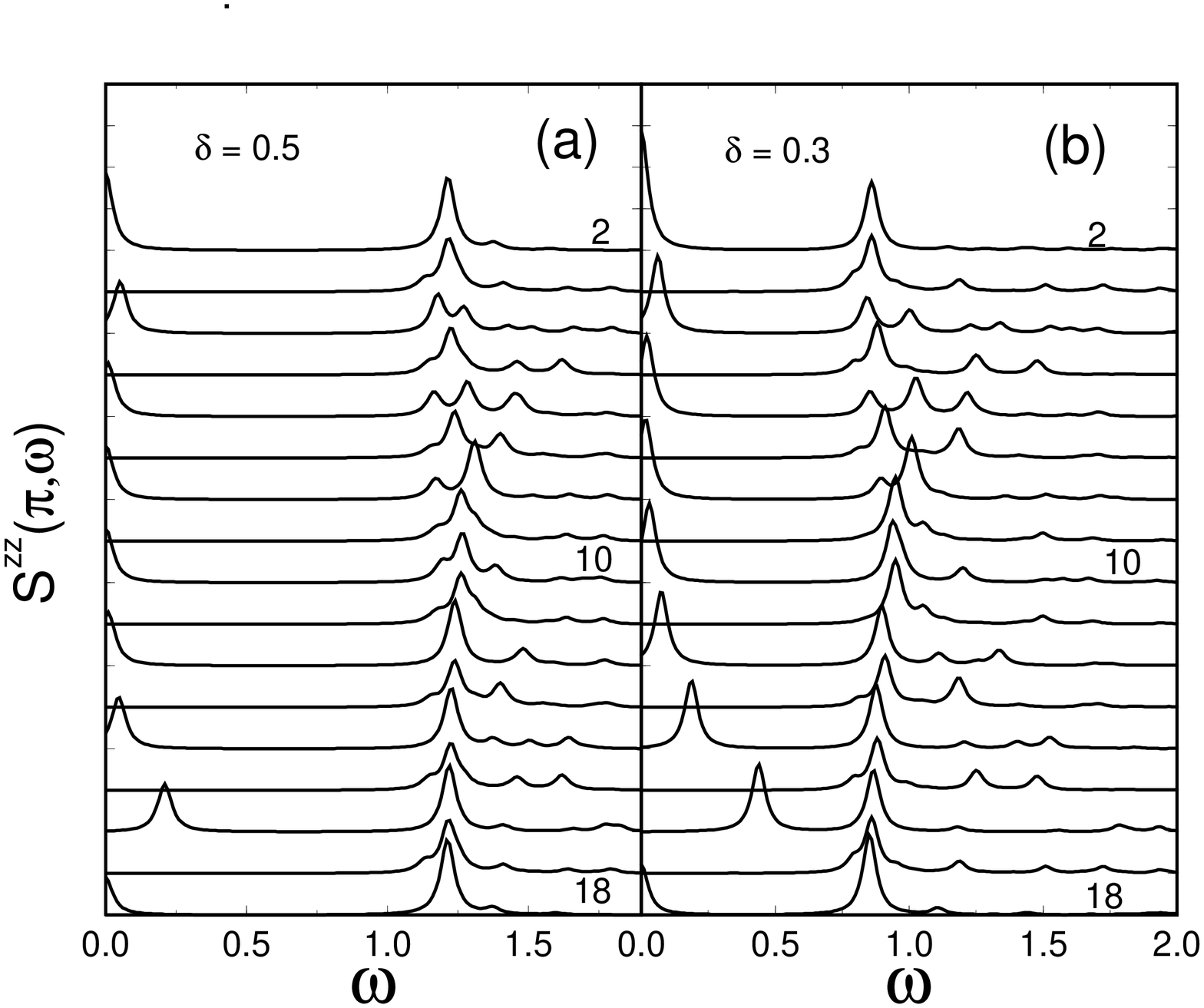,width=9.00cm,angle=-90}
\psfig{figure=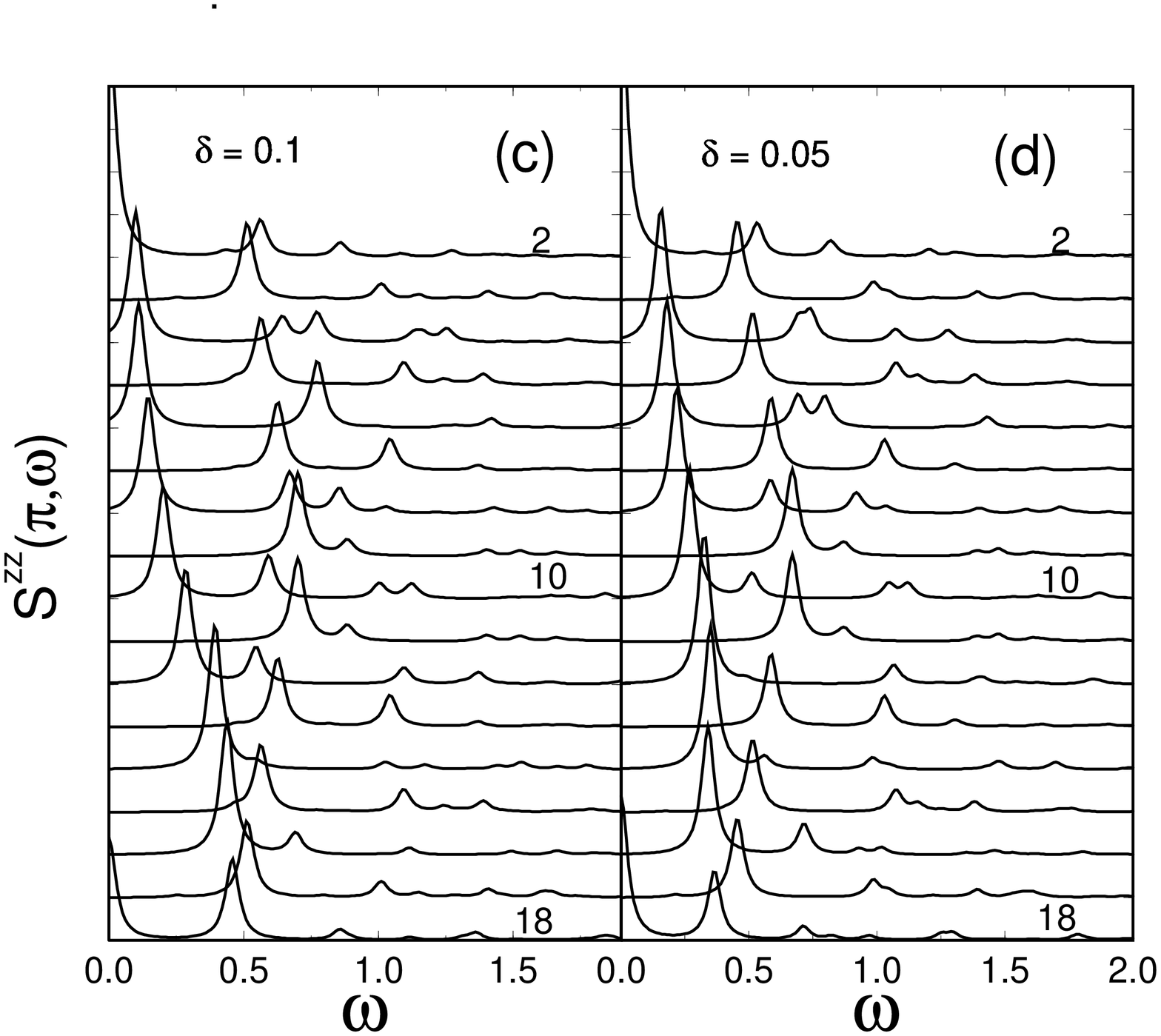,width=9.00cm,angle=-90}
\vspace{0.5cm}
\caption{Dynamical structure factor at $k=\pi$ obtained by exact
diagonalization using a  18 site chain with two $S=1$ impurities,
total $S^z=0$, as a function of the position of the second impurity
and for several values of $\delta$. One of the impurities is
located at site 1, and the position of the second impurity is
indicated increasing from top to bottom.}
\label{szzvsom}
\end{figure}

\noindent
state is zero), and thus the weight of the zero-energy peak cancels. A
similar argument holds for $l_2$ equal to any odd-integer. On the other
hand, for $l_2$
even-integers the matrix elements of the two doped bonds contribute with
the same sign and a low-energy peak should appear in the spectrum.
These results deduced at large $\delta$ actually are qualitatively 
similar when $\delta$ is reduced from 1.0, as it can be observed in
Fig. \ref{szzvsom} where $S^{zz}(\pi,\omega)$ obtained for
$N=18$ is shown. In this figure, it is clear that
even for $\delta$ as small as 0.05 a substantial weight
inside the gap is present only when the second impurity is located in
even-sites, in agreement with the discussion of the previous 
paragraph. In this figure, the
bond between the sites 1 and 2 is ``strong" ($J_i= J(1+\delta)$).
As $\delta$ decreases
the positions of the impurity peaks inside the gap shift to larger
values of $\omega$ and increase their weight, as expected.

\begin{figure}
\hspace{-0.2cm}\psfig{figure=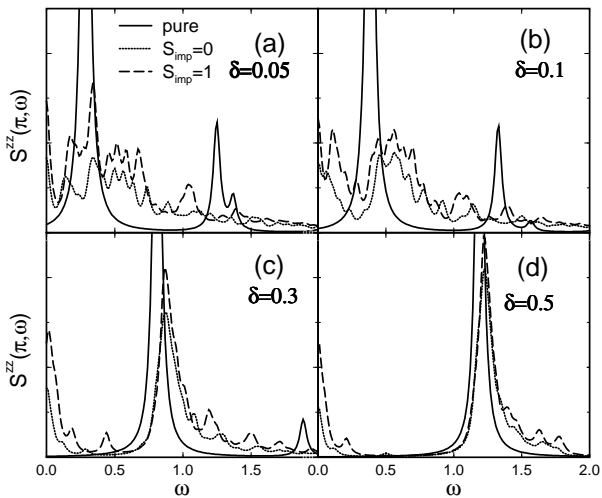,width=9.00cm,angle=-90}
\vspace{0.3cm}
\caption{Dynamical structure factor at $k=\pi$ obtained by exact
diagonalization in the 18 site chain for the pure system, with
two $S=1$ impurities, and with two nonmagnetic impurities,
total $S^z=0$, for several values of $\delta$. These curves have been
obtained by averaging over all impurity positions.}
\label{szzavera}
\end{figure}

In Fig. \ref{szzavera}, $S^{zz}(\pi,\omega)$ for $N=18$
without impurities and in the presence of two $S=1$ impurities for
several values of $\delta$ is presented. For this result an average over
all the impurity
positions has been carried out. In this figure the corresponding
quantity for nonmagnetic impurities is also included
 for comparison. It can be observed
that the weight which appears inside the original gap of the pure 
system is
consistingly larger for $S=1$ impurities than for nonmagnetic
impurities. This weight inside the gap in the case of $S=1$
impurities may correspond to the bound states formed between the
spin-1, considered as two $S=1/2$ spins as in the VBS picture, and
the two neighboring spin-1/2 spins. It is also reasonable to assume
that the weight of these bound states in the ground state is larger
than the weight of the bound states due to the presence of OBC, at
least in the RVB picture of the ground state discussed below, thus
explaining the behavior shown in Fig. \ref{szzavera}.

\vspace{-0.5cm}

\subsection{Enhanced Spin Correlations Near Ni}

It is also  instructive to analyze from a microscopic point of view
the effects of a spin-1 impurity on the local magnetic order. This can
be achieved by studying the spin-spin correlations around the
impurity defined as $S(l)=<S^z(i_1)S^z(i_1+l)>$, where $i_1$ is the site
next to the impurity located at $i_0$, and $l \ll N$. 
A normalization such that  $S(0)=1$ was adopted.
To predict  the behavior of this correlation
it is possible to invoke the short-range RVB picture of
the ground state of spin-1/2 1D chains previously used to describe the 
effect of nonmagnetic impurities~\cite{laukamp,martins1}. In this 
picture, the ground state is assumed to be dominated 
by the superposition of the two possible
dimerized states, one with singlets on the odd bonds and the other
with singlets on the even bonds. 
In the case of nonmagnetic impurities, a simple argument shows that the 
correlation $S(l)$ is {\it enhanced} near the impurities with
respect to the same correlations for the pure system. The idea is simply
that without vacancies a given spin spends roughly half the time coupled
into singlets with each of its two neighbors, while when the spin being
studied is located next to a vacancy only one neighbor is available 
for singlet formation and naturally the correlation between these two
spins is enhanced. The argument can be extended to distances larger than
one lattice spacing easily~\cite{martins2}.

By using this same RVB picture for the case of spin-1 impurities one
obtains, however, a somewhat different behavior.
In the first place let us
observe that the energy of a dimer formed by a spin-1 and a
spin-1/2 is $-J_i$, whereas the energy of a ``pure" dimer
made out of two spin-1/2
has energy $-3J_i/4$. Then,  
the spin-1/2 next to the impurity should have a stronger tendency
to form a singlet with the impurity rather than with other spin-1/2
sites, simply to minimize the energy. 
In addition, and most importantly, is that the ground
state energy of the three site cluster formed by the impurity and
its two spin-1/2 neighbors is equal to ($-2J$), independent of
$\delta$. 
If one considers a subsystem with four sites containing the
$S=1$ impurity, it can be deduced that in order  to minimize the energy
this subsystem prefers to
form a three-site cluster leaving a spin-1/2 unpaired  rather than
two dimers, one of them containing the impurity.  This is true 
at least for $\delta < 1/7$ if the two dimers are formed on
``strong'' bonds, and for all $\delta > 0$ if they correspond to
``weak" bonds. Thus, it is expected that states containing the ground
state of this three-site 1/2-1-1/2 system, which corresponds to a singlet,
plus an unpaired spin and a RVB configuration for the rest of the chain
will have a large overlap
with the actual ground state of the problem.
The coefficients of the states in the three-site system ground state
favors the AF correlations between the three sites. These two
factors imply that: 1. there is a strong AF correlation between a 
site next to the impurity and the impurity itself, $S_{imp}(1)$, 2. 
there is an
{\it enhancement} of the  correlation between both sites next to
the impurity, $S_{cross}(2)$, and 3. there is a
{\it reduction} of the correlations between site $i_1$ (the site next
to the impurity) and the rest of the spins located on the same side 
with respect to the impurity. In summary, what is predicted is a very 
local enhancement of the correlations around the spin-1 impurity. 
The region involved in the process corresponds to only 3 spins 
(the spin-1 and the two neighboring spin-1/2). From the
point of view of the rest of the spins on the chain, this three-spin
structure has formed a singlet and it behaves $effectively$ as a
nonmagnetic vacancy. 
Thus, this result for dimerized chains agrees with a general result
obtained for uniform AF Heisenberg chains which states that an
impurity with spin-S is equivalent to open boundary conditions
with a decoupled spin of size (S-1).\cite{eggert1}
Hence, a spin-1 impurity
is equivalent to a non-magnetic impurity and forms a singlet.

These features can be clearly seen in Table \ref{table1}.
The correlations $S(l)$ have been computed with a standard world-line
Monte Carlo at a temperature $T=0.0625 J$ and for $N=80$, except
otherwise stated. The dimerization is $\delta=0.05$.
The statistical error of the QMC simulations affects only the last 
digit of the quantities reported.
The effects of nonmagnetic impurities are shown for the 40 site
chains for the case where OBC are used. Results for the 39 site
chains are very similar to the values for the 40 site chain.
The effects of $S=1$ impurities are shown in the
last four rows. In these cases the position of the second impurity 
$i_0$
is shown (the first one is located at site 1). In all cases, it
has been indicated if the bond between the reference site, $i_1$, and
its spin-1/2 nearest neighbor site is strong (s)  or weak (w).
In Table \ref{table2},
the dependence of the previous results with $\delta$ is studied 
using Lanczos diagonalization results for a chain with
$N=20$ sites and one impurity. Note the fairly good agreement 
between these zero temperature results compared with the finite
temperature QMC results on the 80 site chain both for the pure
system and for the case of two impurities located at sites $i_0=1, 41$
and $\delta=0.05$.
It can be observed on  Table \ref{table2} that as $\delta$ is
increased from 0.05 to 0.5 there is a smooth variation in the
behavior of the various correlations shown. In all cases $S(1)$,
$S(2)$ and $S(3)$ are reduced from their pure values, and
$S_{imp}(1)$ and $S_{cross}(2)$ are enhanced.

The possibility raised above that the singlet ground state of the 
three-site 
subsystem 1/2-1-1/2 could behave as an effective nonmagnetic 
vacancy, and hence that it may lead to an enhancement of the
AF correlations~\cite{martins2,laukamp} between the sites close
to this vacancy, is actually confirmed by the
results shown in Table \ref{table3}. In this Table, the correlations
between a spin next to one of these three-site clusters, i.e. located
at two lattice spacings from the $S=1$ impurity, and the remaining 
spins of the chain are shown to be {\it enhanced} with respect to 
their values in the pure system. As previously 
noticed~\cite{martins2,laukamp}, the enhancement
of the correlations for the case when the bond between the two first
sites next to a vacancy is weak is more pronounced than the case
of a strong bond between them.

\begin{table}
\caption{Spin-spin correlations $S(l)$ at various distances as defined in
the text obtained by Monte Carlo simulations on chains with $N=80$ sites,
using periodic
boundary conditions, and at $T=0.0625$ unless otherwise stated. 
``Pure'' means that there are no impurities. The
chains with $N=39, 40$ sites have open boundary conditions, and the
correlations are measured starting at the site next to the boundary. The
dimerization is $\delta=0.05$. The last four rows correspond to the
chain in the presence of two $S=1$ impurities, the first one located
at site 1 and the second one at site $i_0$. The reference site is
located at $i_1$. $w,s$ refer to weak and strong bonds, respectively,
between the spin at site $i_1$ and its spin-1/2 neighbor.}
\vspace{0.1cm}
\begin{tabular}{|c|c|c|c|c|c|}
    &  S(1)& S(2) & S(3)  & S$_{cross}(2) $ & S$_{imp}(1) $  \\  \hline
pure s         & -0.777 & 0.230 & -0.247 &  ---  &  ---   \\ \hline
pure s (T=0.05)& -0.785 & 0.230 & -0.250 &  ---  &  ---   \\ \hline
pure w         & -0.400 & 0.230 & -0.122 &  ---  &  ---   \\ \hline
pure w (T=0.05)& -0.408 & 0.230 & -0.126 &  ---  &  ---   \\ \hline
N=40 s         & -0.906 & 0.242 & -0.274 &  ---  &  ---   \\ \hline
N=40 s (T=0.05)& -0.907 & 0.244 & -0.275 &  ---  &  ---   \\ \hline
N=40 w         & -0.777 & 0.297 & -0.253 &  ---  &  ---   \\ \hline
N=40 w (T=0.05)& -0.777 & 0.299 & -0.254 &  ---  &  ---   \\ \hline
imp. i$_0$=40 
i$_1$=41 s     & -0.408 & 0.170 & -0.097 & 0.295 & -0.971 \\ \hline
imp. i$_0$=40 
i$_1$=39 w     & -0.313 & 0.189 & -0.081 & 0.295 & -1.067 \\ \hline
imp. i$_0$=41 
i$_1$=40 s     & -0.429 & 0.196 & -0.141 & 0.293 & -1.148 \\ \hline
imp. i$_0$=41
i$_1$=42 w     & -0.293 & 0.193 & -0.087 & 0.293 & -1.240 \\
\end{tabular}
\label{table1}
\end{table}

\begin{table}
\caption{Spin-spin correlations at various distances as defined in the
text obtained by Lanczos diagonalization for $N=20$, using periodic
boundary conditions and for several values of the dimerization 
$\delta$.}
\vspace{0.1cm}
\begin{tabular}{|c|c|c|c|c|c|}
    &  S(1) & S(2) & S(3)  & S$_{cross}(2) $ & S$_{imp}(1) $ \\  \hline
pure w ($\delta=0.05$) & -0.4113 & 0.2319 & -0.1339 & --- & --- \\ \hline
pure w ($\delta=0.1$)  & -0.3213 & 0.2111 & -0.0927 & --- & --- \\ \hline
pure w ($\delta=0.3$)  & -0.1706 & 0.1401 & -0.0329 & --- & --- \\ \hline
pure w ($\delta=0.5$)  & -0.0954 & 0.0860 & -0.0117 & --- & --- \\ \hline
pure s ($\delta=0.05$) & -0.7679 & 0.2319 & -0.2466 & --- & --- \\ \hline
pure s ($\delta=0.1$)  & -0.8459 & 0.2111 & -0.2378 & --- & --- \\ \hline
pure s ($\delta=0.3$)  & -0.9503 & 0.1401 & -0.1574 & --- & --- \\ \hline
pure s ($\delta=0.5$)  & -0.9834 & 0.0860 & -0.0931 & --- & --- \\ \hline
 w ($\delta=0.05$) & -0.3012 & 0.1953 & -0.0976 & 0.2936 & -1.2385 \\ \hline
 w ($\delta=0.1$)  & -0.2579 & 0.1828 & -0.0760 & 0.2838 & -1.2600 \\ \hline
 w ($\delta=0.3$)  & -0.1511 & 0.1272 & -0.0300 & 0.2179 & -1.3018 \\ \hline
 w ($\delta=0.5$)  & -0.0870 & 0.0795 & -0.0110 & 0.1408 & -1.3211 \\ \hline
 s ($\delta=0.05$) & -0.4325 & 0.1951 & -0.1438 & 0.2936 & -1.1462 \\ \hline
 s ($\delta=0.1$)  & -0.5135 & 0.1832 & -0.1569 & 0.2838 & -1.0723 \\ \hline
 s ($\delta=0.3$)  & -0.7789 & 0.1315 & -0.1396 & 0.2179 & -0.7346 \\ \hline
 s ($\delta=0.5$)  & -0.9206 & 0.0841 & -0.0901 & 0.1408 & -0.4342 \\
\end{tabular}
\label{table2}
\end{table}

\begin{table}
\caption{Spin-spin correlations $S(2,l) = \langle S^z(2) S^z(2+l)
\rangle $ at various distances from
a site two lattice spacings apart from a $S=1$ impurity obtained by
Lanczos diagonalization for $N=20$ for various values of $\delta$,
and by
Monte Carlo simulations for $N=80$ sites and $\delta=0.05$.}
\vspace{0.1cm}
\begin{tabular}{|c|c|c|c|c|}
                &   S(2,3) &  S(2,4) &   S(2,5) &  S(2,6) \\ \hline
 pure s ($\delta=0.05$) & -0.7679 & 0.2319 & -0.2466 & 0.1243 \\ \hline
 pure s ($\delta=0.10$) & -0.8459 & 0.2111 & -0.2378 & 0.0960 \\ \hline
 pure s ($\delta=0.30$) & -0.9503 & 0.1401 & -0.1574 & 0.0360 \\ \hline
 pure s ($\delta=0.50$) & -0.9834 & 0.0860 & -0.0931 & 0.0125 \\ \hline
      s ($\delta=0.05$) & -0.8224 & 0.2371 & -0.2580 & 0.1255 \\ \hline
      s ($\delta=0.10$) & -0.8696 & 0.2141 & -0.2421 & 0.0976 \\ \hline
      s ($\delta=0.30$) & -0.9531 & 0.1404 & -0.1577 & 0.0361 \\ \hline
      s ($\delta=0.50$) & -0.9838 & 0.0861 & -0.0931 & 0.0125 \\ \hline
 pure w ($\delta=0.05$) & -0.4113 & 0.2319 & -0.1339 & 0.1243 \\ \hline
 pure w ($\delta=0.10$) & -0.3213 & 0.2111 & -0.0927 & 0.0960 \\ \hline
 pure w ($\delta=0.30$) & -0.1706 & 0.1401 & -0.0329 & 0.0360 \\ \hline
 pure w ($\delta=0.50$) & -0.0954 & 0.0860 & -0.0117 & 0.0125 \\ \hline
      w ($\delta=0.05$) & -0.6263 & 0.2708 & -0.2113 & 0.1602 \\ \hline
      w ($\delta=0.10$) & -0.5069 & 0.2633 & -0.1606 & 0.1417 \\ \hline
      w ($\delta=0.30$) & -0.2183 & 0.1678 & -0.0464 & 0.0492 \\ \hline
      w ($\delta=0.50$) & -0.1040 & 0.0926 & -0.0133 & 0.0141 \\ \hline
 N=80 pure s      & -0.777  & 0.230  & -0.247  & 0.115 \\  \hline
 N=80 pure w      & -0.400  & 0.230  & -0.122  & 0.116 \\  \hline
 N=80 i$_0$=40 s  & -0.815  & 0.231  & -0.251  & 0.117 \\ \hline
 N=80 i$_0$=40 w  & -0.619  & 0.260  & -0.200  & 0.143 \\ \hline
 N=80 i$_0$=41 s  & -0.829  & 0.235  & -0.260  & 0.120 \\ \hline
 N=80 i$_0$=41 w  & -0.629  & 0.271  & -0.210  & 0.157 \\
\end{tabular}
\label{table3}
\begin{figure}
\psfig{figure=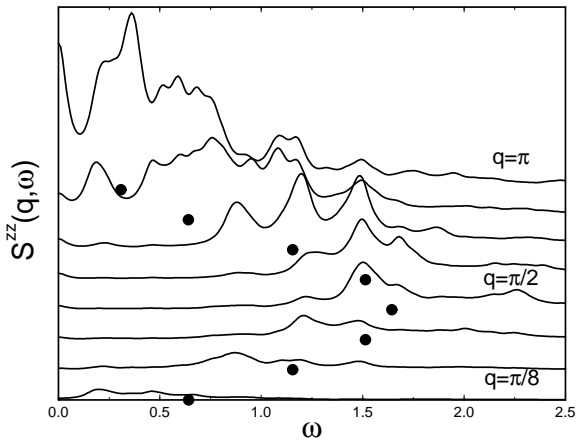,width=9.00cm,angle=-90}
\vspace{0.1cm}
\caption{Dynamical structure factor $S^{zz}(q,\omega)$ obtained by exact
diagonalization in the 16 site chain with two $S=1$ impurities 
for $\delta=0.05$ averaged over all the impurity positions. The
energy of the first peak for the pure $N=16$ chain are indicated
with full circles.}
\label{szzvsk}
\end{figure}
\end{table}

In summary, the results shown thus far seem to indicate that
for magnetic impurities the AF correlations between these
impurities and their two neighboring spin-1/2's are strong.
However, note that once the three-spin
cluster centered at the impurity is considered as a singlet, then the
rest of the spins will enhance the correlations among themselves as if
in the presence of a nonmagnetic impurity.

Finally, in Fig.\ref{szzvsk} $S^{zz}(k,\omega)$ is shown for $N=16$,
$\delta=0.05$ and in the presence of two $S=1$ impurities. The results
are averaged over all the impurity positions. The energies of the first
peaks for the pure system have been indicated for reference. The 
behavior of $S^{zz}(k,\omega)$ in the presence of $S=1$ impurities is 
very similar to the previously reported behavior for nonmagnetic
impurities~\cite{martins1}. In other words, two bands appear here:
1. the low energy ``impurity band" with the highest intensity peaks at 
or near $k=\pi$, and 2. a high energy band, which is essentially a 
remnant of the dispersion relation for the undoped system.

\section{Magnetic susceptibility at finite temperatures}
\label{monte}

The suppression of the spin gap due to magnetic impurities can also
be detected in thermodynamical properties such as the magnetic
susceptibility $\chi(\rm T)$. It is well-known that the presence
of a spin gap between the singlet ground state and the excited
triplet state leads to an exponential decay as the temperature is
reduced to zero. On the other hand, $\chi(\rm T)$ would diverge
or converge to a nonzero value as T vanishes for a gapless system,
such as the uniform 1D Heisenberg model. A similar
result should emerge as a consequence of the presence of states
inside the original gap induced by impurities, as discussed in the
previous section. This change of the magnetic susceptibility upon
Zn, Mg or Ni doping was indeed observed experimentally in the SP
compound CuGeO$_3$.\cite{lussier,hase2,masuda}
In particular, in susceptibility measurements along the c axis
it has been reported the presence of a second maximum at 
temperatures lower than the SP transition 
temperature.\cite{lussier,masuda}
This second maximum in the susceptibility along the c axis has been 
attributed to the transition to a long-range AF order
and it occurs at a temperature slightly higher than the
corresponding Ne\'el temperature (T$_{\rm N}$). 

\begin{figure}
\psfig{figure=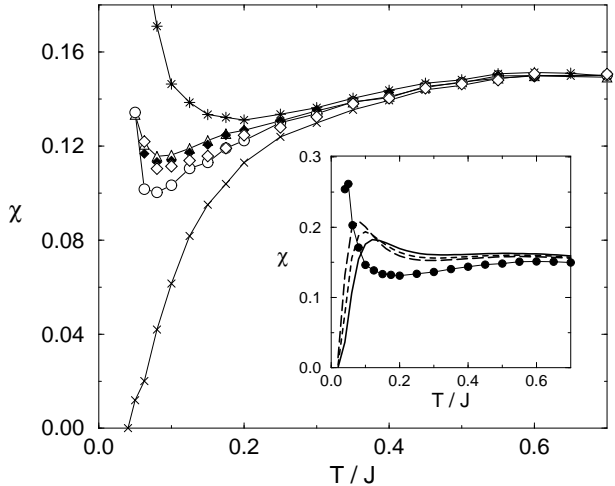,width=9.00cm,angle=-90}
\vspace{0.4cm}
\caption{Magnetic susceptibility as a function of temperature
obtained by Monte Carlo simulations in dimerized Heisenberg chains
with $\delta=0.05$. $N=80$, two $S=1$ impurities in sites 1 and 40 
(up triangles), in sites 1 and 41 (open circles),
in sites 1 and 20 (filled diamonds), and in sites 1 and 21
(open diamonds). $N=40$, OBC, weak end bonds (stars), strong
end bonds (crosses). In the inset, it is shown $\chi(\rm T)$
obtained by exact diagonalization in chains of $N=12$ (solid line),
$N=14$ (short dashed line), and $N=16$ (long dashed line), and
OBC with weak end bonds. The results obtained by QMC for $N=40$, OBC, 
weak end bonds (circles) for temperatures down to $\rm T/J = 0.04$
are also shown.
}
\label{suscep}
\end{figure}

In Fig.\ref{suscep}, the magnetic susceptibility computed
by QMC simulations is shown as a function of temperature for a
dimerized Heisenberg chain with $\delta=0.05$ in the presence of both
magnetic and non-magnetic impurities.
The statistical error of the QMC simulations is about the size 
of the open circles in this figure. The
susceptibility for the pure system on a chain of 80 sites has also
been computed. The
results for this case are very similar to the case of $N=40$ with OBC
and strong edge bonds
and they are not included in the figure for clarity.
$\chi(\rm T)$ for $N=80$ with two spin-1
impurities has also been computed, for the case where the first impurity
is located at site 1 and the second one
at various sites such as 20, 21, 40 and 41. The
susceptibilities obtained for these systems are compared 
with those obtained from
chains of $N=40$ with open boundary conditions. 
The latter chain can be considered as
resulting from a $N=80$ chain after a pair of nonmagnetic impurities
have been introduced. 

In the case of $N=40$, OBC, two
possibilities have been also analyzed: 
the case where the bonds at the chain ends are both
weak and the case where they are both strong. The results
can be interpreted in terms of the intuitive picture
developed in Ref. \onlinecite{martins1} for the large
dimerization limit. In the case of strong end links,
the spins are basically paired in strong dimers and
hence it is quite likely that a spin gap is still present
in the chain leading to an exponential decay of $\chi(\rm T)$
as $T \rightarrow 0$ observed in the figure. This behavior is
not observed in the other case of weak end bonds where there are 
loose spins at the ends
of the chain which cause a strong suppression of the spin gap.
In the case of spin-1 impurities, depending on their relative
positions, a periodic chain is effectively
cut, or in chains with odd number of sites, or in chains with
even number of sites where
at least one of them has both weak end bonds. In both cases, the
susceptibility would not decay exponentially to zero as the
temperature goes to zero as it is observed in the figure at least down
to the lowest temperature achievable with our algorithm.

In principle, the experimentally observed behavior of the 
susceptibility should be theoretically accounted for by averaging the
results obtained for chains of even number of sites with the ones
obtained for chains with odd number of sites. The latter
diverge as $T \rightarrow 0$ due to the total $S^z=1/2$.
For the case of nonmagnetic impurities (open boundary conditions),
from both QMC and exact diagonalization results for chains up to
16 sites, we conclude that by averaging the susceptibilities of 
chains with even number of sites and {\em strong} end bonds with the 
susceptibilities of chains with odd number of sites, one does not
obtain a second maximum in the low temperature region. 
The Curie behavior of the odd chains dominates over the 
exponentially decaying behavior of the even chains at low 
temperatures. On the other
hand, the susceptibility on chains with even number of sites and
{\em weak} end bonds, as it can be seen in the
inset of Fig.\ref{suscep}, does indeed show this second maximum
and it is quite likely that this second maximum can survive
the averaging procedure as we have verified numerically.

Let us examine to what extent these results are relevant to
spin-Peierls systems. In these systems, upon the introduction of 
nonmagnetic
impurities, the spin-lattice coupling leads to the case of chains
where the two edge bonds are strong, even for an odd total
number of sites in which case a soliton is formed in the middle of the
chain.\cite{hansen} However, it has been recently claimed that 
elastic interchain
interactions may stabilize a given pattern of distortions, i.e. with
and without nonmagnetic impurities the weak and strong bonds of the
dimerized chain may remain the same as for non-dynamical
phonons~\cite{khom} and this behavior has been verified 
numerically.\cite{hansen} That is, even site chains with both
weak links are still present in SP chains due to elastic coupling
to pure neighbor chains.

Besides, it is reasonable to think that the results obtained
for larger open chains are going to be indicative of smaller
nonmagnetic impurity doping. Then, as it can be seen in the 
inset of Fig.\ref{suscep}, the shift to lower temperatures
of the second peak as $N$ is increased is in complete agreement
with the experimental results showing that the Ne\'el temperature
{\em decreases} as the impurity concentration 
decreases\cite{masuda}. One should notice
that this behavior of T$_{\rm N}$ occurs in the dimerized region
of CuGeO$_3$ where precisely our fixed-dimerization study should
apply. (A different behavior in the uniform region for large impurity
concentrations has been reported\cite{masuda}). Finally, the fact
that the second maximum in $\chi$ is due to the contribution of
even site chains with weak end links is very consistent with the fact
that, as mentioned above , these chains present the largest 
enhancement of AF correlations upon doping.

On the other hand, the susceptibility in the case of two magnetic
impurities has already a very similar temperature dependence for 
different
impurity positions and it is quite likely that an average over all
these impurity positions would still result in a curve of the same 
shape. Based on the mapping of this problem to an effective nonmagnetic
impurity chain a second maximum should exist in $\chi$ and it
should appear at temperatures lower than
the attainable ones within our QMC algorithm.

\section{Conclusions}
\label{conclu}

In this article  the effects of $S=1$ magnetic
impurities on dimerized AF Heisenberg chains have been analyzed, 
and the results have been compared
with those originated by the introduction of nonmagnetic
impurities. The overall effects, such as the destruction
of the spin gap which reveals itself in the dynamical spin structure 
factor $S(q,\omega)$
(Figs. \ref{szzavera} and \ref{szzvsk}) and in the magnetic
susceptibility (Fig. \ref{suscep}), are similar between the two types 
of impurities. However, the microscopic origin and
details of the tight states formed between the impurities and
the spin-1/2 sites close to it are different. This difference
translates into the existence of strong AF correlations
in the immediate vicinity of the $S=1$ impurity.
This local behavior could eventually be experimentally
detected, for example by using NMR techniques. 
On the other hand, 
the spins located at distances of two lattice spacings (or larger) from
the impurity react to the spin $singlet$  state arising from the
coupling of the S=1
impurity with its two neighboring spin-1/2 as if it were a
nonmagnetic impurity.
This interesting effect produces an
enhancement of AF correlations in the vicinity of the $S=1$ impurities
as for Zn doping, which may  lead to a global stabilization of
the AF state as found experimentally  in
CuGeO$_3$ using several techniques.

As discussed in the previous section with respect to the case of
nonmagnetic impurities, it is expected that in spin-Peierls systems
the partial substitution of Cu ions by Ni ions would also change the
dimerization pattern which these compounds present below the SP
transition temperature. That is, the changes in the spin-spin 
correlations discussed in Section \ref{lanczos} would imply, due to the
spin-phonon coupling, a distortion of the dimerization pattern, 
especially near the Ni impurities. Hence our results could in 
principle not be directly related to the experimentally observed
behavior of Ni-doped CuGeO$_3$. 
However, some previous studies, as well as new results which we
have recently obtained,
indicate that the elastic coupling between a doped chain with
its neighbor undoped ones tends to stabilize the undistorted
dimerized pattern and in this case our results could
be directly relevant to experiments in SP systems. In this sense,
the agreement of our results for the magnetic susceptibility
with recent experimental results on CuGeO$_3$ is remarkable
and this leads to an
interesting explanation on the origin of the second peak
observed in these experiments. On the other
hand, the effects of $S=1$ impurities analyzed in this paper could be
also detected in the structurally
dimerized compounds such as the ones indicated in the Introduction.

\section{acknowledgments}

E. D. thanks the financial support of the NSF grant
DMR-9520776. J. R. wishes to thank D. Poilblanc for many useful
discussions.


\begin{references}

\bibitem{haldane} F. D. M. Haldane, Phys. Rev. Lett. {\bf 50},
        1153 (1983).

\bibitem{renard} J. P. Renard, {\it et al.}, Europhys. Lett. {\bf 73},
       3463 (1994).

\bibitem{ladders} For a recent review see E. Dagotto and T. M. Rice,
    Science {\bf 271}, 618 (1996). See also {\it Physics Today},
     Search and Discovery, pg. 17 October 1996.

\bibitem{azuma} M. Azuma, Z. Hiroi, M. Takano, K. Ishida, Y. Kitaoka,
      Phys. Rev. Lett. {\bf 73}, 3463 (1994).

\bibitem{majumdar} C. K. Majumdar and D. K. Ghosh, J. Math. Phys.
      {\bf 10}, 1388 (1969).

\bibitem{hase} M. Hase, I. Terasaki and K. Uchinokura, Phys. Rev. 
        Lett. {\bf 70}, 3651 (1993).

\bibitem{rieradobry} J. Riera and A. Dobry, Phys. Rev. B {\bf 51}, 
        16098 (1995).

\bibitem{castilla} G. Castilla, S. Chakravarty and V.J. Emery, Phys. 
        Rev. Lett. {\bf 75}, 1823 (1995).

\bibitem{bonner} See J. C. Bonner, S. A. Friedberg, H. Kobayashi,
       D. L. Meier, and H. W. J. Bl\"ote, Phys. Rev. B {\bf 27},
       248 (1983), and references therein.

\bibitem{lake} B. Lake, D. A. Tennant, R. A. Cowley, J. D. Axe,
       and C. K. Chen, J. Phys. Cond. Mat. {\bf 8}, 8613 (1996),
       and references therein.

\bibitem{hiroi} Z. Hiroi, S. Amelinckx, G. Van Tendeloo, and N.
       Kobayashi, Phys. Rev. B {\bf 54}, 15849 (1996).

\bibitem{garrett} A. W. Garrett, S. E. Nagler, D. A. Tennant, B.
       C. Sales, and T. Barnes, Phys. Rev. Lett {\bf 79}, 745
       (1997).

\bibitem{ajiro} See for example, Y. Ajiro {\it et al.}, Phys. Rev. B
      {\bf 51}, 9399 (1995).

\bibitem{lussier} J. G. Lussier, S. M. Coad, D. F. McMorrow, and D.
      McK. Paul, J. Phys. Cond. Matt. {\bf 7}, L325 (1995).

\bibitem{sasago} Y. Sasago, N. Koide, K. Uchinokura, M. C. Martin, M. 
      Hase, K. Hirota, and G. Shirane, preprint cond-mat/9603185, 1996.

\bibitem{hase2} M. Hase, I. Terasaki, Y. Sasago, K. Uchinokura, and H.
    Obara, Phys. Rev. Lett. {\bf 71}, 4059 (1993).

\bibitem{oseroff} S. B. Oseroff, S-W. Cheong, B. Atkas, M. F. Hundley,
    Z. Fisk, and L. W. Rupp, Jr., Phys. Rev. Lett. {\bf 74}, 1450
    (1995).

\bibitem{luca} A. K. Hassan, L. A. Pardi, G. B.  Martins, G. Cao,
       and L. C. Brunel, preprint cond-mat/9706286.

\bibitem{eggert1} S. Eggert and I. Affleck, Phys. Rev. B {\bf 46}, 10866
       (1992).

\bibitem{eggert2} S. Eggert and I. Affleck, Phys. Rev. Lett. {\bf 75},
      934 (1995).

\bibitem{martins2} G. B. Martins, M. Laukamp, J. Riera, and E. Dagotto,
       Phys. Rev. Lett. {\bf 78}, 3563 (1997).

\bibitem{laukamp} M. Laukamp, G. B. Martins, C. J. Gazza, A. L. Malvezzi,
       E. Dagotto, P. M. Hansen, A. C. L\'opez, and J. Riera, 
      Phys. Rev. B {\bf 57}, 10755 (1998).

\bibitem{sandvik} A. Sandvik, E. Dagotto, and D. Scalapino, 
       Phys. Rev. B {\bf 56}, 11701 (1997).

\bibitem{fukuyama} H. Fukuyama, T. Tanimoto, and M. Saito, J. Phys. Soc. 
        Jpn. {\bf 65}, 1182 (1996);
        H. Fukuyama, N. Nagaosa, M. Saito, and T. Tanimoto,
        J. Phys. Soc. Jpn. {\bf 65}, 2377 (1996); 
        N. Nagaosa, A. Furusaki, M. Sigrist, and H. Fukuyama,
        J. Phys. Soc. Jpn. {\bf 65}, 3724 (1996).

\bibitem{martins1} G. B. Martins, E. Dagotto, and J. Riera, 
       Phys. Rev. B {\bf 54}, 16032 (1996).


\bibitem{azuma2} M. Azuma, M. Takano, and R. S. Eccleston, preprint
       cond-mat/9706170. See also M. C. Martin, M. Hase, K. Hirota, G.
       Shirane, Y. Sasago, N. Koide, and K. Uchinokura, preprint.

\bibitem{lemmens} P. Lemmens, M. Fisher, G. G\"untherodt, C. Gros, P.
       G. J. van Dongen, M. Weiden, W. Richter, C. Geibel,  and F.
       Steglich, preprint cond-mat/9703060.

\bibitem{ladderexper}
    M. Azuma, Y. Fujishiro, M. Takano, T.~Ishida, K.~Okuda, 
    M.~Nohara, and H.~Takagi, Phys. Rev. B {\bf 55}, R8658 (1997); 
    M.~Nohara, H.~Takagi, M.~Azuma,
    Y. Fujishiro, and M. Takano, preprint.

\bibitem{ladderall}
   Y. Motome, N. Katoh, N. Furukawa, and M. Imada, J. Phys. Soc. Jpn. 
    {\bf 65}, 1949 (1996);
    M. Sigrist and A. Furusaki, J. Phys. Soc. Jpn. {\bf 65}, 2385 (1996);
     H.-J. Mikeska, U. Neugebauer, and U. Schollw"ock, preprint
       cond-mat/9608100;
      T. K. Ng, Phys. Rev. B {\bf 54}, 11921 (1996);
        T. K. Ng, preprint cond-mat/9610016;
       Y. Iino and M. Imada, J. Phys. Soc. Jpn. {\bf 65}, 3728 (1996);
      M. Imada and Y. Iino, J. Phys. Soc. Jpn. {\bf 65}, 568 (1997);
      K. Hida, preprint cond-mat/9612232;
     T. Miyazaki, M. Troyer, M. Ogata, K. Ueda, and D. Yoshioka, preprint
     cond-mat/9706123; M. Fabrizio and R. M\'elin, preprint 
      cond-mat/9703102; and references therein.

\bibitem{s1chains} 
    S. Miyashita and S. Yamamoto, Phys. Rev. B {\bf 48}, 913 (1993);
      Erik Sorensen and Ian Affleck, Phys. Rev. B {\bf 49}, 15771
      (1994); and references therein.

\bibitem{fabrizio} M. Fabrizio and R. M\'elin, preprint cond-mat/9706158,
      1997.

\bibitem{s1exp} G. E. Granroth {\it et al.}, preprint cond-mat/9710161,
        1997.

\bibitem{affleck} I. Affleck {\it et al.}, Phys. Rev. Lett. {\bf 59},
       799 (1987).

\bibitem{anderson} P. W. Anderson, Mater. Res. Bull. {\bf 8}, 153 (1973).

\bibitem{kivelson} S.A. Kivelson, D.S. Rokhsar, and J.P. Sethna,
       Phys. Rev. B {\bf 35}, 8865 (1987).

\bibitem{augier} D. Augier, D. Poilblanc, E. Sorensen, I. Affleck,
      preprint cond-mat/9802053.

\bibitem{feiguin} A. E. Feiguin, J. A. Riera, A. Dobry, H. A. Ceccatto,
     Phys. Rev. B {\bf 57}, 14607 (1997).

\bibitem{hansen} P. Hansen, D. Augier, J. Riera, and  D. Poilblanc,
	    preprint cond-mat/9805325.

\bibitem{review} E. Dagotto, Rev. Mod. Phys. {\bf 66}, 763 (1994).

\bibitem{nio} E. Dagotto, J. Riera, A. Sandvik, and A. Moreo, 
       Phys. Rev. Lett. {\bf 76}, 1731 (1996).  

\bibitem{sugai} S. Sugai et al., Phys. Rev. B {\bf 42}, 1045 (1990).

\bibitem{WLMC} J. E. Hirsch {\it et al.}, Phys. Rev. B {\bf 26},
         5033 (1982).

\bibitem{cobalt} P. E. Anderson, J. Z. Liu, and R. N. Shelton, 
        Phys. Rev. B {\bf 56}, 11014 (1997).

\bibitem{masuda} T. Masuda, {\it et al.}, Phys. Rev. Lett. {\bf 80}, 
  18 May (1998); K. Manabe, {\it et al.}, preprint cond/mat/9805072.

\bibitem{khom} D. Khomskii, W. Geertsma, and M. Mostovoy, Czechoslovak
  Journal of Physics, vol. 46 (1996), Suppl. S6 (Proceedings of the 21st
  International Conference on Low Temperature Physics, Prague, August 
  1996).

\end{references}
\end{document}